\documentclass[fleqn,twoside,twocolumn,nofootinbib,showkeys]{revtex4} 
\usepackage[sec,nocpr]{ujp} 
\usepackage{multirow}

\begin{document}
\title[Frequency dependence of conductivity]
{FREQUENCY DEPENDENCE OF CONDUCTIVITY IN THE LAYERED I\MakeLowercase{N}$_4$S\MakeLowercase{E}$_3$ CRYSTALS}%
\author{Stakhira J. M.}
\affiliation{Ivan Franko University of Lviv}
\address{50, Dragomanov Str., Lviv 79005, Ukraine}
\author{Demkiv T. M.}
\affiliation{Ivan Franko University of Lviv}
\address{50, Dragomanov Str., Lviv 79005, Ukraine}
\author{Fl'unt O. Y\MakeLowercase{e}.}
\affiliation{Ivan Franko University of Lviv}
\address{50, Dragomanov Str., Lviv 79005, Ukraine}

\udk{621.325.592}  \razd{\secvii}

\autorcol{J.M.\hspace*{0.7mm}Stakhira, T.M.\hspace*{0.7mm}Demkiv, O.Y\MakeLowercase{e}.\hspace*{0.7mm}Fl'unt}%

\setcounter{page}{737}%

\begin{abstract}
The frequency dependence of conductivity in In$_4$Se$ _3$, pure and with copper admixture crystals, in the region of nitrogen temperatures is investigated. It was found out that variable length hops on the localized levels in the vicinity of Fermi level is pre-dominant mechanism of charge transfer in the crystals. For pure In$ _4$Se$ _3$ crystals the density of localized states is $10^{17}-10^{18}$ eV$^{-1}\cdot$cm$^{-3}$, the mean hop length is 220--350 {\AA}. The reasons for the occurrence of localized states are considered within the model of a layered crystal as a quasi-disordered system.\end{abstract}


\maketitle

\section*{Introduction}
One of the possible mechanism of charge transport in layered crystals is thermally activated electron hops on the deep traps in the forbidden band, for example, GaSe[1], InSe[2], In$_4$Se$_3$[3]. Results of works [1--3] indicates that considering the temperature dependence of conductivity in layered crystals it's necessary to take into account singularities of crystal structure of these semiconductors. Particularly, decreasing of the density of localized states near Fermi level in In$_4$Se$_3$[3] at uniaxial mechanical strain along a normal to the layers is connected with existence of quasi-localized states, the nature of which is connected with vibrational movement of the layers on the electronic subsystem of the crystal.

In this paper the result of measurement of frequency dependence of complex conductivity in a layer (100) of pure and copper doped In$_4$Se$_3$ crystals has been presented.

\section{Experimental methods}

Single crystals of In$_4$Se$_3$ were grown by Chochralski method with using of Peltier effect with following thermal annealing. Entered into crystals copper is donor type and decreases conductivity by a few orders of magnitude. Electrical contacts to sample were made by method described in [4]. Amplitude of applied ac electrical field corresponds to linear range of voltage-current characteristic. Measurement of ac electrical conductivity 
was carried out by transforming of complex impedance into proportional ac voltage [5, p. 130--131]. The main element of the equipment was transformer of complex impedance $Z_x$ of sample under study on the base of operational amplifier (OA). Back feed loop contains reference resistance $R_0$. Complex transfer coefficient within the working region for ideal OA depends only on $Z_x$.This allows to avoid the effect of stray capacitors on the result of measurement. Real part of complex conductivity (later conductivity) $\sigma$ on certain frequency can be obtained from the expression
\[
\sigma=k\sin \varphi / R_0\:, 
\]
where $k$ -- gain module of the transformer and $\varphi$--phase shift between output and input signals that are measured during experiment. Taking into account true amplitude-phase characteristic of the real OA using developed computer program decreases possible relative measurement error down to 3\% for $\sigma\approx10^{-2}-10^{-5}$ Ohm$^{-1}$ and  5\% for $\sigma\approx10^{-6}-10^{-8}$ Ohm$^{-1}$.

\section{Results and discussion}
As it can be seen from Fig. 1 on frequencies above characteristic frequency $\omega_0\approx$ 500 Hz the conductivity $\sigma(\omega)$ becomes frequency dependent and has complex form. Two linear dependencies are observed in log-log scale with significantly different slopes. Experimental dependencies well fit by math expression $\sigma=\sigma(0)+A_1\omega^{s_1}+A_2\omega^{s_2}$, where $A_1$, $A_2$, $s_1$ and $s_2$ -- constants, $\sigma(0)$ -- dc conductivity.

\begin{figure}
\vskip1mm
\includegraphics[width=\column]{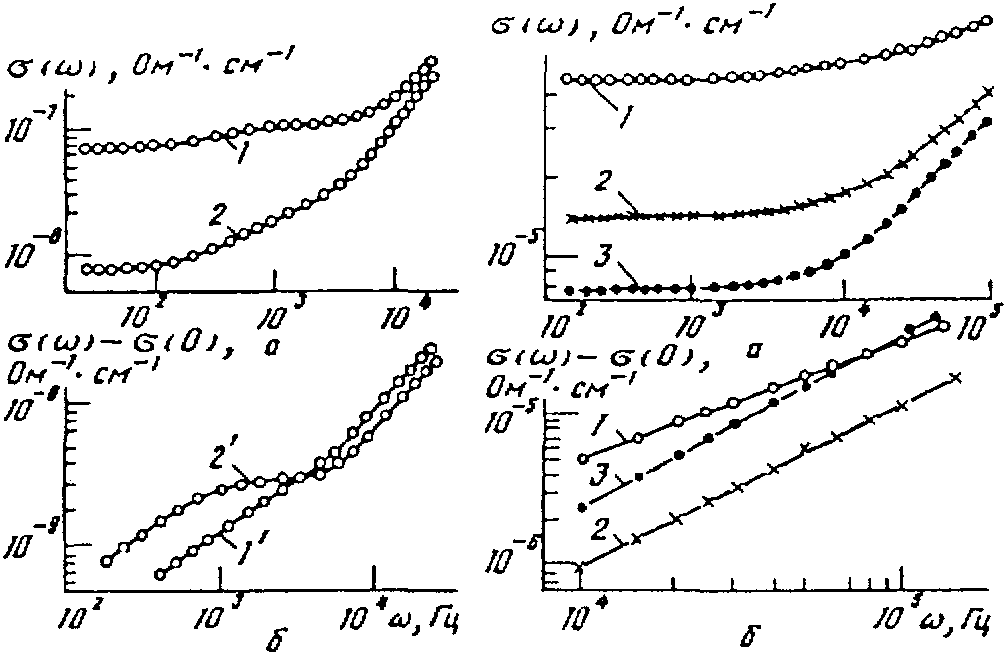}
\vskip-3mm\caption{Frequency dependence of conductivity $\sigma(\omega)$ (a) and $\sigma(\omega)-\sigma(0)$ (b) for current direction along [001] in In$_4$Se$_3$ crystals at 125 K (1, 1') and 175 K (2, 2')}
\end{figure}

\begin{figure}
\vskip-3mm\caption{The same as on fig. 1 for In$_4$Se$_3$ crystals with 1.2~at.~\% of Cu at 225 K (1), 175 K (2) and 125 K (3)}
\end{figure}

Frequency dependent conductivity described above with temperature-independent coefficient $s=0.7-0.83$ is typical for variable range hopping conductivity via narrow band of localized levels in the vicinity of Fermi level and can be described by Austin-Mott model [6, 7]. According to this model frequency dependence of electrical conductivity and exponent $s$ are described by expressions
\[
\sigma(\omega)=1/96\pi^3e^2kTN_F^2\alpha^{-5}\omega[\ln(\omega_{ph}/\omega)]^4\:,
\]
\[
s=1-4/\ln(\omega_{ph}/\omega)\:,
\]
where $N_F$ -- density of localization levels in the vicinity of Fermi level, $\alpha$ -- wave function decay constant, $a=1/\alpha$ -- radius of localization, $\omega_{ph}$ -- characteristic phonon frequency accompanying localized electron hops.

\begin{table}[b]
\noindent\caption{}\vskip3mm\tabcolsep4.5pt

\noindent{\footnotesize
\begin{tabular}{|c|c|c|c|c|}
 \hline%
 \rule{0pt}{3mm}
 
 \multirow{2}{*}{Parameter}
 & \multicolumn{2}{|c}{$s_1$}
 & \multicolumn{2}{|c|}{$s_2$}\\
 
 \cline{2-5}
 \rule{0pt}{3mm}
 &125 K&175 K&125 K&175 K\\
 
\hline%
\rule{0pt}{5mm}

$s$&0.7&0.7&0.81&0.81\\

$\omega_{ph}$, Hz&$1\cdot10^9$&$6\cdot10^8$&$1\cdot10^{13}$&$1\cdot10^{13}$\\

$N_{F}$, eV$^{-1}\cdot$cm$^{-3}$&$3\cdot10^{17}$&$7\cdot10^{17}$&$1\cdot10^{17}$&$2\cdot10^{17}$\\

$N_{t}$, cm$^{-3}$&$4\cdot10^{16}$&$8\cdot10^{16}$&$8\cdot10^{15}$&$2\cdot10^{16}$\\

$R_{c\omega}$, {\AA}&220&220&350&350\\

$\Delta\varepsilon$, eV&0.03&0.03&0.02&0.02\\
\hline
\end{tabular}
}
\end{table}
According to theory of hopping ac conductivity average life time of charge carriers $\tau$ with adsorption or realizing of phonon was obtained from the expression
\[
\tau^{-1}=\omega_{ph}\exp(2\pi R_{c\omega})\:,
\]
where $R_{c\omega}$ -- average hop length. 
\[
(kT)^3/N_FR_{c\omega}^3\Delta\varepsilon^4=1
\]
Thereby, localized near Fermi level states that give the contribution to conductivity are stricted within the energy band $\Delta\varepsilon$ nearby the Fermi level. It gives possibility to estimate the concentration of them using the formula $N_t=\Delta \varepsilon N_F$.

Results of calculation show (see Table 1), that characteristic phonon frequency $\omega_{ph}$ for low-frequency range is by three orders of magnitudes lower, than for high-frequency one. Value of $\omega_{ph}$ for high-frequency range corresponds to atoms vibration frequencies in In$_4$Se$_3$ [9]. Also the concentration of localized levels $N_t$ with $\omega_{ph}$ is almost by one order of magnitude higher than concentration of localized states with $\omega_{ph}$ that corresponds to frequencies of individual atoms in indium subselenide. So the conducted measurements indicate that charge carrier's transport in high-quality In$_4$Se$_3$ crystals at nitrogen temperatures occurs by hops over localized levels of two different types.

Hops with $\omega_{ph}=10^{13}$ Hz are characteristic for charge hops via localized levels caused by additional atoms. Hops with characteristic phonon frequency about $10^9$ Hz take place over localized levels the nature of which is connected with singularities of crystal structure of layered crystals, that leads to existence of collective electron-phonon states [8, 9]. Temperature dependence of frequency $\omega_{ph}$ and concentration of charge carriers for low frequency range confirms the applicability of this approach. 

Within the frame of considered model it could be explained logically an absence of additional linear in log-log scale frequency dependence for samples In$_4$Se$_3$ with Cu and In additives (fig. 2a). Critical frequency $\omega_{0}$ is about $10^3$ Hz. On frequencies higher than $\omega_0$ conductivity obeys the law $\sigma\sim\omega^{s}$. Temperature sensitivity of exponent $s$ indicates that tunneling probability of charge carriers in additional band is not negligibly small. Thereby the charge transport in In$_4$Se$_3$ crystals occurs mainly by hopping than tunneling across potential barriers dividing the localized states.

\section*{Conclusion}
So the character of frequency dependence of copper doped In$_4$Se$_3$ single crystals indicates that the variable range hopping on localized states near Fermi level conductivity is dominant mechanism of electrical charge transfer at low temperatures. Drastic simplifying of the frequency dependence of conductivity in copper doped In$_4$Se$_3$ single crystals can be explained by decreasing of intensity of vibrational movement of layers in the result of trimerization of the crystals [8, 9].

\vspace*{-5mm}\rezume{
Й. М. Стахіра, Т. М. Демків,  О. Є. Флюнт}{ЧАСТОТНА ЗАЛЕЖНІСТЬ ПРОВІДНОСТІ У ШАРУВАТИХ КРИСТАЛАХ I\MakeLowercase{N}$_4$S\MakeLowercase{E}$_3$} {Досліджено частотну залежність провідності в шаруватих кристалах In$_4$Se$_3$, чистих та з домішками міді, в області азотних температур. Виявлено, що переважаючим є механізм перенесення заряду зі змінною довжиною стрибка по локалізованих рівнях в околі рівня Фермі. Для нелегованих кристалів In$_4$Se$_3$ густина локалізованих станів становить $10^{17}-10^{18}$ eV$^{-1}\cdot$cm$^{-3}$, середня довжина стрибка -- 220--350 {\AA}.
Причини виникнення локалізованих станів розглядаються в рамках моделі шаруватого кристала як квазірозупорядкованої системи.}

\end{document}